\documentclass[11pt]{article}
\usepackage{graphicx}  % standard LaTeX graphics tool
\usepackage{amssymb,latexsym}%,amsmath}
\usepackage{ntheorem}
\newcommand{\bea}{\begin{eqnarray}}
\newcommand{\eea}{\end{eqnarray}}
\usepackage[mathscr]{eucal}
\usepackage{citesort}

\newtheorem{Theorem} {\sc  Theorem\rm} [section]
\newtheorem{Remark}[Theorem]{\sc  Remark\rm}

\newcommand{\qed}{\hfill $\Box$ \medskip}
\begin{document}

\title{Conditions for nonexistence of static or stationary,
  Einstein-Maxwell, non-inheriting black-holes}

\author{Paul Tod\\Mathematical Institute\\Oxford}

\maketitle
\begin{abstract}

We consider asymptotically-flat, static and stationary solutions of
the Einstein equations representing Einstein-Maxwell space-times in which
the Maxwell field is not constant along the Killing vector defining
stationarity, so that the symmetry of the space-time is not
inherited by the electromagnetic field. We find that static
degenerate black hole solutions are not possible and, subject to
stronger assumptions, nor are static, non-degenerate or stationary
black holes. We describe the possibilities if the stronger
assumptions are relaxed.
\end{abstract}

\newcommand{\beq}{\begin{equation}}
\newcommand{\eeq}{\end{equation}}
\newcommand{\zA}{\mathring{A}}
\newcommand{\zX}{\mathring{X}}%
\newcommand{\mcL}{\mathcal{L}}%

\section{Introduction}
\label{sI}
In the study of static or stationary Einstein-Maxwell solutions of
Einstein's equations, it is frequently assumed that the Maxwell field is
also static or stationary, in the sense that the Lie derivative of the
Maxwell tensor $F_{ab}$ along the Killing vector $K^a$ is zero. One needs the
energy-momentum tensor $T_{ab}$ to be static or stationary for the Einstein
equations to be consistent, but this does not actually require that $F_{ab}$ be
static or stationary. Following custom, (see e.g. \cite{MV}) we shall say that $F_{ab}$ does not {\emph{inherit}} the
symmetry if $T_{ab}$ is static or stationary but $F_{ab}$ is not.

It is not always assumed that $F_{ab}$ is static or stationary, and
there are explicit non-inheriting solutions in the literature (see
e.g. \cite{MV}, \cite{es}). However none of these solutions is
asymptotically-flat and it may be wondered whether the properties of
being asymptotically-flat and non-inheriting are incompatible. As we
shall see, this is certainly the case for Maxwell fields in
Minkowski space, and it is very likely though not proven to be the
case for static space-times without horizons. However, in the
presence of a black hole, it is easier to proceed by a study of the
horizon rather than a study of the asymptotic conditions. We
consider static and stationary, asymptotically-flat, Einstein-Maxwell 
black-hole solutions, where a black-hole solution is understood to
be an asymptotically-flat solution with a regular Killing horizon
bounding the domain of outer communication (d.o.c.). We shall assume
 that black holes are (topologically) spherical, and we conclude that,
subject to conditions we give, indeed there are none which are
non-inheriting. (In \cite{cw} it is shown that black holes are necessarily topologically spherical if certain regularity conditions are satisfied; by adding these conditions to our assumptions, we could omit the separate assumption of sphericity.)

 The first result is the following:
\begin{Theorem}
\label{11} Let $(M,{}^4g)$ be an asymptotically-flat, Einstein-Maxwell,
space-time with a Killing vector $K^a$ which is time-like near
infinity.

 Then
necessarily
\beq
\mcL_KF_{ab}=-aF^*_{ab},\label{kv0}
\eeq
for some $a$, where $\mcL_K$ is the Lie derivative along $K^a$ and
$F^*_{ab}$ is the dual of $F_{ab}$. We shall suppose that the
right-hand-side in (\ref{kv0}) is not
everywhere zero, and that any black holes are (topologically) spherical. then:

\begin{enumerate}
\item
If $M$ is static, then $a$ is constant and there are
no black holes with degenerate horizons.
\item
If $M$ is static, then there are no black holes for which $M$ is analytic
up to and including the horizons.
\item
If $M$ is stationary with the Killing vector in (\ref{kv0}) tangent
to the generators of the horizon and \emph{either} non-null Maxwell
field \emph{or} null Maxwell field with constant $a$ then there are
no black holes for which $M$ is analytic up to and including the
horizons.
\item
If $a$ is a non-zero constant, then the charge of any black hole
vanishes.
\end{enumerate}

\end{Theorem}

It is not assumed here that the solutions contain just a single
black hole, nor that the horizons are nondegenerate, nor that the
Maxwell field is non-null (though, as we shall see, this is
necessarily true for the static case). The assumption in part 3,
that the Killing vector which is time-like at large distances is
also tangent to the generator of the horizon might be thought
unreasonable, but we shall see from the proof that this part still
holds in an analytic solution of the kind envisaged, provided the
null generator of the horizon satisfies (\ref{kv0}) with non-null
Maxwell field or with null Maxwell field but a nonzero constant $a$.
% We shall discuss the analyticity condition below, but it is worth
% observing that a static or stationary Einstein-Maxwell solution will
% necessarily be analytic (in suitable coordinates){\bf Bit more here} where the Killing
% vector is time-like. In particular therefore, if $a$ is constant or
% the Maxwell field is null in an open set of the region in which the
% Killing vector is time-like, the same will hold in the connected
% component of this region.

One
usually expects a
stationary black-hole solution also to be axisymmetric and one could
then
imagine both symmetries not being inherited. Part 3
will apply provided the Killing vector generating the horizon is not
inherited, but in the span of two non-inherited symmetries there will
be an inherited one and this could be the null generator of the
horizon (see the end of Section 3 for this argument). That this can't happen in
the orthogonally-transitive case
with a non-null Maxwell field
follows from a result of Michalski and
Wainwright \cite{MW}:

\medskip

{\emph{If $(M,{}^4g)$ is a stationary, axisymmetric Einstein-Maxwell space-time which is orthogonally-transitive with a non-null Maxwell
field, then necessarily the
Maxwell field inherits the symmetry in that
\[\mcL_KF_{ab}=\mcL_LF_{ab}=0,\]
where $K$ and $L$ are the stationarity and axisymmetry Killing
vectors.}

\medskip

There is no assumption of asymptotic flatness or of analyticity in
this result. For vacuum or inheriting Einstein-Maxwell space-times,
orthogonal transitivity can be deduced from conditions we have
imposed (see e.g. \cite{he}), but this does not seem to be the case
for non-inheriting Einstein-Maxwell where it seems to be a definite
restriction. The above result of \cite{MW} assumes that the Maxwell
field is non-null in order to use the `already-unified' formalism
\cite{MisWhe} and in particular the `complexion' vector field, whose
definition we shall recall in Section 3. With the aid of this
formalism one can show the following:
\begin{Theorem}
\label{12}
 Let $(M,{}^4g)$ be an Einstein-Maxwell space-time having a non-degenerate
Killing horizon generated by a Killing vector $K^a$ with
\[
\mcL_KF_{ab}=-aF^*_{ab},\]
and subject to the regularity conditions of \cite{RW}, so that,
without loss of generality, there is a bifurcation surface on the
Killing horizon. Suppose that $F_{ab}$ is nowhere null in the d.o.c. and that the
complexion vector field $\alpha_a$ extends smoothly to the horizon,
then $a=0$ in the set ${\mathcal{B}}$ which is the union of the
d.o.c. and the horizon.
\end{Theorem}
Thus for a non-null Maxwell field and a non-degenerate horizon, if
the complexion vector field extends smoothly to the horizon, one can
dispense with the assumption of analyticity up to the horizon. In
particular, this would show that there are no non-degenerate
non-inheriting static black holes, even without this assumption of
analyticity, since for the static case the Maxwell field is obliged
to be non-null. This with Part 1 of Theorem~\ref{11} would rule out
static black holes, but the assumption that $\alpha_a$ extends
smoothly to the horizon is a strong one.

To prove Theorem 1.2, following the argument of \cite{MW}, one first
establishes that $K^a\alpha_a=2a$ which is constant. We do this in
Section 3. Then, with the regularity assumptions of \cite{RW}, which
are very natural in this context, the non-degeneracy of the Killing
horizon implies that, without loss of generality, there is a
bifurcation surface on the Killing horizon where $K^a$ vanishes.
Therefore $a=0$, and the symmetry is inherited. (The same argument
incidentally shows that the symmetry is inherited if $K^a$ tends to
a constant time-translation at large distances while $\alpha_a$
decays to zero.)

In the stationary but non-static case, with null fields,
Theorem~\ref{11} rules out constant, nonzero $a$ with analyticity at
the horizon. We also have the following:

\begin{Theorem}
\label{13}
 Let $(M,{}^4g)$ be an Einstein-Maxwell space-time, with a
Killing horizon generated by a Killing vector $K^a$ with
\[
\mcL_KF_{ab}=-aF^*_{ab},\]
and a null Maxwell field. Then \emph{either} $a$ is constant
\emph{or} the space-time admits a non-twisting, shear-free null geodesic congruence.

\end{Theorem}

If there exists a non-twisting, shear-free null geodesic congruence, then the metric lies in the Robinson-Trautman class if the expansion is non-zero, or the Kundt class if the expansion is zero, or is a pp-wave if the generator of the congruence can be chosen to be parallel. It is very unlikely that a metric in either of the last two classes could be stationary and asymptotically-flat, as the curvature does not decay along the congruence. In the Robinson-Trautman class, the metric can be given locally in terms of a few
functions, with (in our case) one remaining field equation (see
Theorem 28.3 of \cite{es} for this). Explicit non-inheriting metrics
can be found (see e.g. the metric of Bartrum \cite{bart} given in
\cite{es}) but again it seems rather unlikely (as we argue in the
Appendix) that these metrics, with the Maxwell field null but
non-zero, can be stationary and asymptotically-flat. 

\begin{Remark}\label{14}

 \rm{A final point worth making is that, even for an inherited
symmetry, at a non-degenerate horizon a null Maxwell field vanishes
to all orders.}
\end{Remark}

\medskip

 The plan of the paper is as follows. In Section 2, we undertake a
 `near horizon' analysis, as in \cite{iwp}, for static, non-inheriting,
 Einstein-Maxwell black holes. We find that the
Maxwell field vanishes at the horizon to all orders. This proves
part 4 of Theorem~\ref{11} in the static case, since the charge of
the black hole is an integral of the Maxwell field over a
cross-section of the horizon. For a static, degenerate horizon, the
analysis of \cite{iwp} together with the vanishing of the Maxwell
field at the horizon leads to a contradiction, so that at once there
are no non-inheriting solutions. This proves part 1 of
Theorem~\ref{11}.

For a (static or stationary) non-degenerate horizon or a stationary
degenerate horizon we need something more, and we assume analyticity
up to and including the horizon. This is a strong 
assumption: there are general arguments that static or stationary
vacuum solutions are analytic away from the
horizon, \cite{MH}, \cite{mh2}, and the
black hole uniqueness theorems have always led to solutions which
are analytic at the horizon but, aside from \cite{chr} for static
vacuum black holes, there are no direct proofs of analyticity up to
and including the horizon. With this assumption we are done: the
Maxwell field vanishes to all orders at the horizon so by
analyticity it vanishes everywhere and we are back to vacuum black
holes - there are no static non-inheriting black holes, which is
part 2 of Theorem~\ref{11}.

In Section 3, we look at stationary, non-inheriting black-hole
space-times. A near-horizon analysis again shows that, for non-null
Maxwell field or null Maxwell field with constant $a$, the Maxwell
field vanishes at the horizon to all orders, and then the assumption
of analyticity forces it to vanish everywhere and we have proved the
slightly stronger version of part 3 of Theorem~\ref{11}. This also
proves part 4 for stationary solutions. For Theorem~\ref{12} we
review the relevant parts of the already-unified theory of
\cite{MisWhe}. For Theorem~\ref{13} we take the analysis of null
Maxwell fields further. The repeated spinor of the Maxwell spinor is
necessarily geodesic and shear-free, and one is able to conclude
that $a$ is constant unless the spinor field is also non-twisting,
which characterises the Robinson-Trautman solutions. We conclude by
noting the remaining loop-holes for non-inheriting black hole
solutions.

In the Appendix we argue, without a complete proof, that there are
in fact no Robinson-Trautman solutions with a (non-zero) null
Maxwell field which are stationary and asymptotically-flat.

\medbreak

For the rest of this section, we develop some theory of
non-inheriting solutions.

We assume that there is a Maxwell field determined by a bivector
field $F_{ab}$ given in terms of a symmetric spinor field
$\phi_{AB}$ by
\beq\label{f0}
F_{ab}=\phi_{AB}\epsilon_{A'B'}+\overline{\phi}_{A'B'}\epsilon_{AB},
\eeq
The corresponding energy-momentum tensor can be written
\beq \label{t0} T_{ab}=2\phi_{AB}\overline{\phi}_{A'B'}. \eeq
The dual Maxwell field is given by
\beq\label{fstar}
F^*_{ab}=-i\phi_{AB}\epsilon_{A'B'}+i\overline{\phi}_{A'B'}\epsilon_{AB}.
\eeq
By assumption, we have a Killing vector $K^a$ with $\mcL_K
T_{ab}=0$, where $\mcL_K$ is the Lie-derivative along $K^a$, but the
same does not hold for $F_{ab}$ or $\phi_{AB}$. However, by
(\ref{t0}) we must have
\[\left(\mcL_K\phi_{AB}\right)\overline{\phi}_{A'B'}+\phi_{AB}\left(\mcL_K\overline{\phi}_{A'B'}\right)=0\]
which entails
\beq \label{phi1} \mcL_K\phi_{AB}=ia\phi_{AB} \eeq
for some real $a$ which, at this stage, could vary with position.
From (\ref{phi1}) with (\ref{f0}) and (\ref{fstar}) we obtain
\beq \mcL_K F_{ab}=-aF^*_{ab}\;\;;\;\;\mcL_K
F^*_{ab}=aF_{ab}.\label{f1} \eeq
This is the familiar fact that, if $T_{ab}$ does not change with
time, then $F_{ab}$ is allowed to change only by a duality rotation.

From the exterior derivative of both sides in each of (\ref{f1}),
for source-free fields we obtain
\beq \label{f2} da\wedge F=0=da\wedge F^*. \eeq
With the aid of (\ref{f0}) and (\ref{fstar}), from (\ref{f2}) we
obtain the following equation, written in terms of spinors:
\beq \phi^{AB}\nabla_{AA'}a=0.\label{fn3} \eeq
Multiply this by $\phi_{BC}$ to conclude that for non-null $F_{ab}$,
$da=0$ and so $a$ is constant in the d.o.c. (This goes through for
static \emph{and} stationary space times.) If $a=0$, we are in the
inheriting case, so throughout we shall suppose $a\neq 0$. We cannot
conclude that $a$ is constant by this argument if $F_{ab}$ is null
and will return to this case in Section 3.

Now we restict to the static case so that the time-like Killing
vector $K^a$ is hypersurface-orthogonal. We introduce the usual
coordinates, so that the metric in the d.o.c. (which we assume is simply-connected; this would follow from the assumptions in \cite{cw}) can be written
\beq
\label{met1}
g=V^2dt^2-h_{ij}(x^k)dx^idx^j,
\eeq
where the Killing vector is $K=\partial/\partial t$ and
$g(K,K)=V^2$. Note also that
\beq
K_adx^a=V^2dt
\label{kv1}
\eeq
We shall assume later that $V^2$ has one or more zeroes and that these
correspond to regular Killing horizons.

The surfaces of
constant time in the metric (\ref{met1}) have zero extrinsic curvature
so, from the momentum
constraint for the Einstein equations, one learns at once that
$T_{0i}=0$, or in an invariant formulation
\beq
\label{t1}
T_{ab}K^b=fK_a.
\eeq
Here $f$ is a function on space-time, independent of $t$ by
assumption and non-negative by the Dominant Energy Condition (which is
automatic for an energy-momentum tensor of the form of
(\ref{t0})). In spinors, (\ref{t1}) is
\[2\phi_{AB}\overline{\phi}_{A'B'}K^{BB'}=fK_{AA'}.\]
Multiplying both sides of this by $\phi^{AC}$ we obtain
\[(\phi_{CD}\phi^{CD})\overline{\phi}_{A'B'}K^{B'}_{\;\;A}=-f\phi_{AB}K^B_{\;\;A'},\]
which is only possible if
\bea
\phi_{CD}\phi^{CD}&=&fe^{i\theta}\label{the1}\\
\phi_{AB}K^B_{\;\;A'}&=&-e^{i\theta}\overline{\phi}_{A'B'}K^{B'}_{\;\;A}\label{the2}
\eea
for some real $\theta$. Note that $f=0$ iff $F_{ab}$ is null or zero
(recall that $F_{ab}$ is null iff $\phi_{CD}\phi^{CD}=0$)
and that if $F_{ab}$ is null and nonzero then $K^a$ must be null, so that this can
only happen at a horizon: $F_{ab}$ is non-null in the d.o.c. (The incompatibility of a null Maxwell field
with a static geometry is an old result, to be found in \cite{ban}.)

Now (\ref{phi1}) applied to (\ref{the1}) gives
\beq
\label{the3}
\mcL_K\theta=2a.
\eeq
We shall find below that $\theta=2at$.

Before that, where $V> 0$,  we introduce the electric and magnetic
field vectors by
\[E_a=V^{-1}K^bF_{ba}\;;\;\;\;B_a=V^{-1}K^bF^*_{ba}\]
or by combining them:
\beq
\label{e1}
E_a+iB_a=2V^{-1}K_{A'}^{\;\;C}\phi_{AC}.
\eeq
From this, (\ref{the2}) and algebra, we find
\[E_a\cos(\theta/2)=-B_a\sin(\theta/2)\]
while from (\ref{phi1}) and (\ref{e1}),
\[\mcL_K E_a=-aB_a\;;\;\;\mcL_KB_a=aE_a.\]
We may therefore introduce $W_a$ by
\beq
\label{w1}
E_a=W_a\sin(\theta/2)\;;\;\;B_a=-W_a\cos(\theta/2),
\eeq
to find
\beq
\label{w2}
\mcL_KW_a=0,
\eeq
and, with the aid of (\ref{the1}),
\beq
\label{w3}
W^aW_a=-2f.
\eeq
We can write the Maxwell equations in terms of $W_a$ by first observing
that, since the Killing vector $K^a$ is static
\beq
\nabla_a(V^{-1}K_b)=-V^{-2}K_a\nabla_bV.
\label{kv3}
\eeq
Now from (\ref{f1}),(\ref{kv3}) and the source-free Maxwell equations
we obtain the system of equations:
\begin{eqnarray*}
\nabla^b(V^{-1}E_b)&=&0\\
\nabla^b(V^{-1}B_b)&=&0\\
\epsilon^{ab}_{\;\;\;\;cd}K^d\nabla_a(VE_b)&=&-aVE_c\\
\epsilon^{ab}_{\;\;\;\;cd}K^d\nabla_a(VB_b)&=&-aVB_c.
\end{eqnarray*}
Substituting (\ref{w1}) into these, we find
\bea
\nabla^b(V^{-1}W_b)&=&0\label{w6}\\
\epsilon^{ab}_{\;\;\;\;cd}K^d\nabla_a(VW_b)&=&-aVW_c\label{w7}
\eea
and
\[W^b\partial_b\theta=0=\epsilon^{abcd}K_dW_b\partial_a\theta,\]
which, with (\ref{the3}) implies
\[
\partial_a\theta=2aV^{-2}K_a\]
so that, by (\ref{kv1}) and up to an additive constant,
$\theta=2at$, as anticipated.

In terms of $W_a$, the energy-momentum tensor is
\beq
T_{ab}=-W_aW_b+\frac{1}{2}(W^cW_c)g_{ab}-\frac{1}{V^2}(W^cW_c)K_aK_b.
\label{t2}
\eeq
In terms of the (positive-definite) spatial metric $h_{ij}$, its
Levi-Civita derivative $D_i$, Ricci tensor $R_{ij}$ and Laplacian $\Delta=h^{ij}D_iD_j$, where indices
from this part of the alphabet run from 1 to 3, the Einstein equations are the system:
\bea
\Delta V&=&\frac{1}{2}(h^{ij}W_iW_j)V,\label{v1}\\
\epsilon_i^{\;\;jk}D_j(VW_k)&=&-aW_i,\label{w4}\\
R_{ij}-\frac{1}{2}Rh_{ij}&=&V^{-1}D_iD_jV-W_iW_j,\label{r1}
\eea
From (\ref{w4}) it follows that
\beq
D_iW^i=0\label{w5}
\eeq
and from (\ref{r1}) that the (three-dimensional) Ricci scalar is
\[R=h^{ij}W_iW_j.\]
For asymptotic flatness, at large distances $R=O(r^{-4})$ so that $|W|=O(r^{-2})$ (for
the application following equation (\ref{curl}) below, it would suffice to have
$|W|=O(r^{-\frac{3}{2}-\epsilon})$ or $R=O(r^{-3-\epsilon})$ for $\epsilon>0$).

As a check on these equations, we note that the contracted Bianchi identity:
\[D^i(R_{ij}-\frac{1}{2}Rh_{ij})=0\]
is indeed an identity given (\ref{v1})-(\ref{r1}). Also if $a=0$, so that the symmetry \emph{is}
inherited, then from (\ref{w1}), $VW_i$ is a gradient, say
\[VW_i=\sqrt{2}D_i\phi\]
in terms of a function $\phi$, which, by (\ref{w5}), satisfies the equation
\[D^i(V^{-1}\phi_i)=0.\]
Now this with (\ref{v1}) and
(\ref{r1}) makes up the field equations for an inheriting Einstein-Maxwell
solution given e.g. in \cite{es}.

\medbreak

It is easy to see that these equations are
incompatible with asymptotic flatness in the limit of special
relativity i.e. for a non-inheriting Maxwell field in flat space.
All that remains of the equations in this limit is the Maxwell
equation from (\ref{w4}) which we may write as:
\beq
\nabla\wedge{\bf W}=-a{\bf W}
\label{curl}
\eeq
We may see as follows that this is incompatible with asymptotic flatness in the sense of $|{\bf W}|^2=O(r^{-3-\epsilon})$. Introduce the three-dimensional tensor
\[t_{ij}=W_iW_j-\frac{1}{2}\delta_{ij}|W|^2,\]
then $t_{ij}$ is divergence-free by virtue of (\ref{curl}) so that
\[\partial_i(t^{ij}x_j)=\delta_{ij}t^{ij}=-\frac{1}{2}|W|^2\]
where $x^i$ is the position vector. We integrate this identity over a large
ball and use the divergence theorem and the assumed rate of decay of
$W_i$ to show that the surface term tends to zero, so that the volume integral vanishes and $W_i$ is zero.

The following argument suggests that the same result holds for a static, asymptotically-flat, noninheriting
solution without horizons.
Rescale the metric with the conformal factor $V^{-2}$:
\beq
\tilde{h}_{ij}=V^{-2}h_{ij},\label{h1}
\eeq
so that also
\[\tilde{\epsilon}_i^{\;\;jk}=V\epsilon_i^{\;\;jk},\]
and define $\omega_i=VW_i$. Then (\ref{w4}) becomes
%
%\beq
\[\tilde{\epsilon}_i^{\;\;jk}\partial_j\omega_k=-a\omega_i,\]%\label{h2}
%\eeq
%
or in form notation
\[*\,d\omega=-a\omega.\]
Another derivative of this shows that
\[\Delta\omega:=(*\,d*d+d*d\,*)\omega=a^2\omega,\]
so that $\omega$ is a one-form eigenfunction of the
Laplacian, and the asymptotic conditions imply that $\omega$ is
square-integrable.

It seems very likely (see e.g. footnote 38 on page 76 in
\cite{mel1}) that there are no non-zero square-integrable
eigen-one-forms of the Laplacian for an asymptotically-Euclidean
metric like $\tilde{h}$. If this is so, then $\omega$ is zero and
there are no static, non-inheriting solutions without a black hole.

To deal with black-hole solutions, where $V$ may have zeroes, we turn
to a near-horizon analysis.

\section{Near-horizon analysis}
\label{sNHA} In this section, we assume that we have a static,
non-inheriting Einstein-Maxwell
 black hole with one or more components of the event horizon. All
 black holes have spherical topology. We shall
 find that the Maxwell field vanishes to all orders at the
 horizon. For a degenerate horizon, this is sufficient, with the help
 of results in \cite{iwp} to reach a contradiction. For the
 non-degenerate or stationary cases we shall need an assumption of analyticity.

 As in, for example, \cite{iwp} we introduce Gaussian null
 coordinates near a component ${\cal{N}}$ of the event horizon, so that the metric is
\beq
{}^4g =rA du^2-2dudr-2r(hd\zeta+\overline{h}d\overline{\zeta})du-m\overline{m},
\label{metric1} \eeq
where $m=-\zX d\overline{\zeta}+O(r)$.

In these coordinates, the Killing vector $K$ is $\partial/\partial u$
with norm
\[g(K,K)=rA\]
and ${\cal{N}}$ is located at $r=0$. The surface gravity of the
horizon is $\kappa=\zA$, where the zero
means the value at $r=0$. By a general argument (see e.g.\cite{w} p.333)
$\zA$ is constant on each component of the horizon. If it vanishes,
the horizon is degenerate.

We shall investigate the metric (\ref{metric1}) in the
spin-coefficient formalism \cite{NT}. We introduce the null tetrad
$(l^a,n^a,m^a,\overline{m}^a)$ by
$$
\begin{array}{lllll}
\l^a\partial_a&=&D&=&\partial_u +\frac{rA}{2}\partial_r\;,\\
n^a\partial_a&=&\Delta&=&-\partial_r\;,\\
m^a\partial_a
&=&\delta&=&\frac{1}{\overline{X}}\partial_{\zeta}
+\frac{r}{\overline{Y}}\partial_{\overline{\zeta}}
-\left(\frac{rh}{\overline{X}}+\frac{r\overline{h}}{\overline{Y}}\right)\partial_r\;,
\end{array}
$$
where $X=\zX +O(r)$.

In this tetrad, the Killing vector is given by
\beq\label{kv2}
K^a=l^a+\frac{rA}{2}n^a
\eeq
and it is clear that all elements of the tetrad are Lie-dragged by
$K^a$, whence so is the spinor dyad $(o^A,\iota^A)$ which lies behind
the tetrad. In this dyad, the Maxwell spinor can be expanded in the
standard way as
\[\phi_{AB}=\phi_0\iota_A\iota_B-\phi_1(o_A\iota_B+\iota_A o_B)+\phi_2o_Ao_B,\]
when (\ref{the2}) with (\ref{kv2}) implies
\beq\label{phi3}
\phi_0=\frac{1}{2}rAe^{i\theta}\overline{\phi}_2,
\eeq
while (\ref{phi1}) implies
\beq\label{phi4}
\mcL_K\phi_i=ia\phi_i\eeq
for $i=0,1,2$.

For the spin-coefficients we find
\beq
\epsilon=\frac{1}{4}\zA+O(r),
\label{eps}
\eeq
while $\nu =\kappa=0$, $\sigma$ and $\rho$ are $O(r)$ and the rest
are $O(1)$. We shall need two of the Maxwell equations from \cite{NT}
\bea
D\phi_1-\overline{\delta}\phi_0&=&(\pi-\alpha)\phi_0+2\rho\phi_1-\kappa\phi_2\label{m1}\\
D\phi_2-\overline{\delta}\phi_1&=&-\lambda\phi_0+2\pi\phi_1+(\rho-2\epsilon)\phi_2.\label{m2}
\eea
Consider first (\ref{m1}). We assume that $\phi_1$ and $\phi_2$ are
$O(1)$ then by (\ref{phi3}), $\phi_0$ is $O(r)$. Now all terms on the right are $O(r)$; on the
left however there is an $O(1)$ term $\partial\phi_1/\partial u$ which
is $ia\phi_1$ by (\ref{phi4}). Thus in fact $\phi_1=O(r)$.
Next, from (\ref{m2}) we find
\beq \frac{\partial\phi_2}{\partial
u}+\frac{\zA}{2}\phi_2=O(r),\label{fn2}\eeq
making use of (\ref{eps}). Using (\ref{phi4}) again, since $\zA$ is
real, we find $\phi_2=O(r)$ (for degenerate or non-degenerate horizons).

By (\ref{phi3}) we now have $\phi_0=O(r^2)$ and we can go round
again. By induction, all components $\phi_i$ vanish at the horizon
faster than any positive power of $r$.

For a degenerate horizon, we can refer to the result in \cite{iwp}:
there are no static vacuum solutions with degenerate horizons
subject to regularity conditions being assumed here; now that
$\phi_{AB}$ vanishes at the horizon to all orders, the case under
investigation here has the same equations holding at the horizon as
a vacuum solution and we obtain the same contradiction. Thus there
are no non-inheriting, Einstein-Maxwell black-holes with a degenerate
horizon, which proves part 1 of Theorem~\ref{11}.

For non-degenerate horizons, we need to assume analyticity up to and
at the horizon. Then the vanishing of the Maxwell field to all
orders at the horizon implies that it is everywhere zero and we are
back to the vacuum case, proving part 2 of Theorem~\ref{11}.

\section{The stationary case}
\label{sSC} In this section, we consider the case of stationary,
non-inheriting Einstein-Maxwell black holes. With the Maxwell field as
before, i.e. (\ref{f0}), we obtain (\ref{phi1}), but now we don't
have available the argument that $F_{ab}$ is non-null. If it is then
we can deduce that $a$ is necessarily constant; if it isn't then for
Theorem~\ref{11} we shall just assume that $a$ is constant, and look
later at the case when it isn't. We introduce null Gaussian
coordinates as before and for non-degenerate horizons the
spin-coefficients behave as before (except that now $\kappa$ is
$O(r)$ rather than zero). At this stage in the argument, the Killing
vector $K^a$ is by assumption tangent to the generator of the
horizon.

We don't have (\ref{t1}) and therefore not (\ref{phi3}) either,
which was important in making $\phi_{AB}$ vanish at the horizon.
However there is a general argument (see \cite{w} equation (12.5.2))
that $T_{ab}K^aK^b=0$ on any component of the horizon. This forces
$\phi_0$ to vanish at $r=0$, so that by smoothness $\phi_0=O(r)$.
The argument after (\ref{m1}) and (\ref{m2}) goes through so that
$\phi_1$ and $\phi_2$ are both $O(r)$.

 To go round again we use another Maxwell equation from \cite{NT}:
\beq
\Delta\phi_0-\delta\phi_1=(2\gamma-\mu)\phi_0-2\tau\phi_1+\sigma\phi_2.
\label{m3} \eeq
Every term is known to be $O(r)$ except for $\Delta\phi_0=-\frac{\partial\phi_0}{\partial r}$ so we
conclude that this is too, and then $\phi_0=O(r^2)$. Now we can go round
again to see by induction that $\phi_{AB}$ vanishes to all orders at
the horizon.

For degenerate horizons we find the spin-coefficient $\epsilon$ is
$O(r^2)$ rather than $O(r)$, but with a non-zero constant $a$ the
argument proceeds just as for non-degenerate horizons and the
Maxwell field vanishes to all orders at the horizon.

With the assumption of analyticity up to and including the horizon,
we may conclude that the Maxwell field is zero everywhere: there are
no analytic, non-inheriting stationary black holes with constant
nonzero $a$, which is the third and final part of Theorem~\ref{11}.

\qed

It is convenient here, since the same arguments as above are used,
to prove Remark~\ref{14}, that with a null Maxwell field inheriting
the symmetry any horizon must be degenerate. The condition for a
null Maxwell field is
\beq \phi_0\phi_2=(\phi_1)^2. \label{n1}\eeq
As noted above, at a horizon we have $\phi_0=O(r)$ so that by
(\ref{n1}) $\phi_1$ vanishes at the horizon and then, by smoothness,
is $O(r)$. We obtain (\ref{fn2}) as before but, with the symmetry
being inherited, the first term is zero. Provided the horizon is
non-degenerate, $\zA$ is a nonzero constant so that $\phi_2=O(r)$.
To go round again we use (\ref{m3}) to make $\phi_0$ be $O(r^2)$,
then (\ref{n1}) to find $(\phi_1)^2=O(r^3)$ when smoothness forces
$\phi_1=O(r^2)$. From (\ref{fn2}), $\phi_2=O(r^2)$ and so,
inductively, the Maxwell field vanishes to all orders. We use this
observation in the list of possibilities at the end of Section 3.

\medskip

For Theorem~\ref{12}, we need to review some of the
`already-unified' theory of \cite{MisWhe} (see also section 5.3 of
\cite{PR}; their conventions are slightly modified here for
consistency with our earlier sections). For an Einstein-Maxwell space-time
with an energy-momentum tensor of the form of (\ref{t0}) with a
non-null $\phi_{AB}$, it is possible algebraically to extract from
$T_{ab}$ a symmetric spinor field $\Phi_{AB}$, unique up to sign,
satisfying
\[T_{ab}=2\Phi_{AB}\overline{\Phi}_{A'B'}\]
and
\[\Phi^2:=\Phi^{AB}\Phi_{AB}=\overline{\Phi}^{A'B'}\overline{\Phi}_{A'B'}\geq 0,\]
i.e. $\Phi^2$ is real and positive. This is clear by taking
components in a frame. (The second condition imposed on $\Phi_{AB}$
implies that the corresponding bivector field is simple.) Thus
$T_{ab}$, and therefore by Einstein's equations the Ricci tensor
$R_{ab}$, determine $\phi_{AB}$ up to a phase (since $\phi_{AB}$ and
$\Phi_{AB}$ differ only by a phase). To fix the phase, one
introduces the complexion vector field by
\begin{eqnarray*}
\alpha_a&=&2(R^{pq}R_{pq})^{-1}\epsilon_{abcd}R^{de}\nabla^cR^b_{\;e}\\
&=&2i(\Phi^2)^{-1}(\Phi^B_{\;A}\nabla^C_{\;A'}\Phi_{CB}-\overline{\Phi}^{B'}_{\;A'}\nabla^{C'}_{\;A}\overline{\Phi}_{C'B'}).
\end{eqnarray*}
Note that this is only defined for a non-null $\Phi_{AB}$, which is
why the formalism makes that assumption. If $\phi_{AB}$ satisfies Maxwell's equations then $\alpha_a$ is
necessarily closed as a one-form and then, in a simply-connected
region, the solution of the Maxwell equations is $\phi_{AB}=e^{i\theta/2}\Phi_{AB}$ where
\[\partial_a\theta=\alpha_a.\]
If $K^a$ is a Killing vector then, as in Section~\ref{sI},
necessarily
\[\mcL_K\Phi_{AB}=iA\Phi_{AB}\]
for some real $A$, but now since $\Phi^2$ is everywhere real, $A$
must vanish and so $\Phi_{AB}$ inherits the symmetry. If the Maxwell
field $\phi_{AB}$ does not inherit the symmetry, so that
\[\mcL_K\phi_{AB}=ia\phi_{AB}\]
then, by the argument in Section 1 and the assumption of non-nullness,
$a$ is a constant and
\[\mcL_K\phi^2=2ia\phi^2\]
where $\phi^2:=\phi^{AB}\phi_{AB}$, but then $\phi^2=e^{i\theta}\Phi^2$ so that
\beq
K^a\alpha_a=K^a\partial_a\theta=2a=\mathrm{constant},\label{F}\eeq
which is what is needed for the proof of Theorem~\ref{12}.

\qed

\medskip

For Theorem \ref{13}, suppose we have a null Maxwell field so that
the Maxwell spinor takes the form
\beq \phi_{AB}=\phi_2 o_Ao_B \label{fn1} \eeq
for some function $\phi_2$ and spinor field $o_A$. (We could rescale
the function $\phi_2$ and spinor field $o_A$ so as to preserve
$\phi_{AB}$, even setting $\phi_2=1$ if desired, but it is
convenient to leave it like this.)

As is well-known, and in any case easy to see, the Maxwell equations
force $o_A$ to be geodesic and shear-free.

We obtain (\ref{fn3}) as before but this time this is just
equivalent to
$$o^A\nabla_{AA'}a=0,$$
which doesn't yet force constancy of $a$. We introduce a second
spinor field $\iota_A$ to make a normalised basis with $o_A$, and
define the Newman-Penrose operators $(D,\;\Delta,\;\delta)$ in the
usual way (see e.g. \cite{NT}). Then $a$ satisfies
$$Da=0=\delta a$$
and since $a$ is real, we also have $\overline{\delta}a=0$.

The geodesic shear-free condition implies the vanishing of the
spin-coefficients $\sigma$ and $\kappa$, and then the commutator of
$D$ and $\delta$ on $a$ is \cite{NT}: 
$$(\delta D-D\delta)a=(\rho-\overline\rho)\Delta a=0,$$
making use of $Da=\delta a=\overline\delta a=0$. Therefore, either
$\Delta a$ vanishes, in which case $a$ is constant, or
$\rho-\overline\rho$ vanishes, in which case the congruence is
twist-free. This completes the proof of Theorem~\ref{13}.

\qed

\medskip

As noted above, the Robinson-Trautman (RT) metrics are characterised by
admitting a twist-free, shear-free congruence of null geodesics with $\rho\neq 0$, while for the Kundt class $\rho=0$. The pp-waves have $o^A$ parallel (i.e. covariantly constant). It
is possible to find (local) RT solutions with a null Maxwell field and
a non-inherited symmetry, but, as we describe in the Appendix, it
seems very unlikely that they can be asymptotically-flat. In the other cases, the curvature does not decay in the direction of the congruence so they probably can't be asymptotically-flat either.

\medskip

The arguments of Theorems~\ref{11}-~\ref{13} rule out many
possibilities for stationary, non-inheriting black holes
but the following remain open:
\begin{itemize}
\item
the black hole is static and non-degenerate or stationary, but is
\emph{not} analytic at the horizon (though still with constant $a$
and zero charge); or
\item
the black hole is stationary and axisymmetric with Killing vectors
$K$ and $L$ satisfying
\[\mcL_KF_{ab}=-aF^*_{ab}\;;\;\;\mcL_LF_{ab}=-bF^*_{ab}\;,\]
with $a$ and $b$ nonzero constants, but the null generator of the
horizon is the inherited symmetry $J=bK-aL$, not forbidden by
Theorem~\ref{11}. The Maxwell field must be null: by the argument
leading to (\ref{F}), for a non-null Maxwell field we would have
$b=L^a\alpha_a$ which vanishes at the fixed points of the rotation
$L$, a contradiction. By Remark~\ref{14}, the horizon must be
degenerate. The charge must be zero: since
\[\mcL_L\phi_1=ib\phi_1\]
with $b\neq 0$; now the charge is an integral of $\phi_1$ over the
horizon, so that the angular dependence on $\phi$, where
$L=\partial/\partial\phi$, will force it to vanish; or
\item
the metric lies in the Robinson-Trautman class with the Maxwell
field null, the horizon degenerate, and the generator of the horizon
a non-inherited symmetry with non-constant $a$. (We argue in the
Appendix that this case isn't in fact possible.)
\end{itemize}

\medskip

\section*{Acknowledgements}
This work was prompted by a question asked by Malcolm MacCallum at
the Isaac Newton Institute in Cambridge. I am grateful to him for
the question, to the Isaac Newton Institute and DAMTP, Cambridge,
the Albert Einstein Institute, Golm, and LMPT, Universit\'e de Tours,
for hospitality and to Piotr Chru\'sciel, Mihalis Dafermos and
Richard Melrose for useful discussions, and to an anonymous referee for helpful comments.

\section*{Appendix: Robinson-Trautman metrics with null Maxwell
field} In this appendix, we shall consider Robinson-Trautman metrics
with null Maxwell field and see that they are unlikely to provide
stationary, asymptotically-flat metrics. The discussion is
suggestive rather than conclusive.

A local form for the Einstein-Maxwell Robinson-Trautman metrics is given
in Theorem~28.3 of \cite{es}. To make the Maxwell field null, one
sets $Q=0$, to find the Maxwell field as
$$F=h(u,\zeta)du\wedge d\zeta + \mathrm{c.c.},$$
and the metric (with signature switched to accord with our
conventions) as
$$g=2du(dr+\frac{1}{2}Hdu)-\frac{2r^2}{P^2}d\zeta d\overline{\zeta},$$
where $h(u,\zeta)$ is arbitrary, $P(u,\zeta,\overline{\zeta})$ is
subject to the remaining field equation, $H$ is given by
$$H=\Delta\log P - 2r\frac{P_u}P-2\frac{m}{r}\;,$$
and $\Delta =2P^2\partial_{\zeta}\partial_{\overline{\zeta}}$  (so
$\Delta$ here is \emph{not} the Newman-Penrose operator with that
name).

 It is
hard to see how the Maxwell field $F$ as above can be asymptotically
flat unless $h=0$ : regularity at $\zeta=0$ would require $h$ finite
there, but then regularity at $\zeta=\infty$ would be impossible.
However, one doesn't know exactly how these local coordinates should
be related to coordinates on Scri, so we give a second argument as
follows.

The existence of a Killing vector imposes further restrictions on
the metric: the spinor dyad is fixed implicitly by the equations:
$$o_A\overline{o}_{A'}dx^{AA'}=du\;;o_A\overline{\iota}_{A'}dx^{AA'}=\frac{r}{P}d\zeta,$$
and the Killing vector $K^a$ must preserve $o_A$ up to scale, so
that
$$\mcL_Ko_A=fo_A\;,\mcL_K\iota_A=go_A-f\iota_A\,,$$
for some functions $f$, $g$. We make an ansatz
$$K^a\partial_a=A\partial_u+B\partial_r+C\partial_{\zeta}+
\overline{C}\partial_{\overline{\zeta}}\;.$$
The Killing equations can then be solved to find the only
possibilities as
$$K^a\partial_a=\alpha\partial_u+C(\zeta)\partial_{\zeta}+
\overline{C}(\overline{\zeta})\partial_{\overline{\zeta}}\;,$$
for a constant $\alpha$ and a function $C$ (this form for the
Killing vector can be simplified further, using the allowed
coordinate freedom, but there are still some necessary conditions to
impose).

For the norm of this Killing vector, we find
$$g(K,K)=\alpha^2H-\frac{2r^2}{P^2}C\overline{C}.$$
However, we want the Killing vector to be time-like asymptotically,
which with our signature means $g(K,K)>0$. Assuming, as seems most
likely, that `asymptotically' means `for large $r$', we therefore need $C=0$
(since $H=O(r)$ at most). However $\partial/\partial u$, which is
what then remains, can only be a Killing vector if $\partial P/\partial u
=0$, when (from the remaining field equation) the metric collapses
to the Schwarzschild metric.

\end{document}